# Citation Advantage For OA Self-Archiving Is Independent of Journal Impact Factor, Article Age, and Number of Co-Authors

## Open Access Archivangelism

### Wednesday, January 17. 2007

#### Citation Advantage For OA Self-Archiving Is Independent of Journal Impact Factor, Article Age, and Number of Co-Authors

---

**SUMMARY:** *Eysenbach has suggested that the OA (Green) self-archiving advantage might just be an artifact of potential uncontrolled confounding factors such as article age (older articles may be both more cited and more likely to be self-archived), number of authors (articles with more authors might be more cited and more self-archived), subject matter (the subjects that are cited more, self-archive more), country (same thing), number of authors, citation counts of authors, etc.*

*Chawki Hajjem (doctoral candidate, UQaM) had already shown that the OA advantage was present in all cases when articles were analysed separately by age, subject matter or country. He has now done a multiple regression analysis jointly testing (1) article age, (2) journal impact factor, (3) number of authors, and (4) OA self-archiving as separate factors for 442,750 articles in 576 (biomedical) journals across 11 years, and has shown that each of the four factors contributes an independent, statistically significant increment to the citation counts. The OA-self-archiving advantage remains a robust, independent factor.*

*Having successfully responded to his challenge, we now challenge Eysenbach to demonstrate -- by testing a sufficiently broad and representative sample of journals at all levels of the journal quality, visibility and prestige hierarchy -- that his finding of a citation advantage for Gold OA (articles published OA on the high-profile website of the only journal he tested (PNAS) over Green OA articles in the same journal (self-archived on the author's website) was not just an artifact of having tested only one very high-profile journal.*

---

In May 2006, Eysenbach published "Citation Advantage of Open Access Articles" in PLoS Biology, confirming -- by comparing OA vs. non-OA articles within one hybrid OA/non-OA journal -- the "OA Advantage" (higher citations for OA articles than for non-OA articles) that had previously been demonstrated by comparing OA (self-archived) vs. non-OA articles within non-OA journals.

This new PLoS study was based on a sample of 1492 articles (212 OA, 1280 non-OA) published June-December 2004 in one very high-impact (i.e., high average citation rate) journal: Proceedings of the National Academy of Sciences (PNAS). The findings were useful because not only did they confirm the OA citation advantage, already demonstrated across millions of articles, thousands of journals, and over a dozen subject areas, but they showed that that advantage is already detectable as early as 4 months after publication.

The PLoS study also controlled for a large number of variables that could have contributed to a false OA advantage (for example, if more of the authors that chose to provide OA had happened to be in subject areas that happened to have higher citation counts). Eysenbach's logistic and multiple regression analyses confirmed that this was not the case for any of the potentially confounding variables tested, including the (i) country, (ii) publication count and (iii) citation count of the author and the (iv) subject area and (v) number of co-authors of the article.

However, both the Eysenbach article and the accompanying PLoS editorial, considerably overstated the significance of all the controls that were done, suggesting that (1) the pre-existing evidence, based mainly on OA self-archiving ("green OA") rather than OA publishing ("gold OA"), had not been "solid" but "limited" because it had not controlled for these potential "confounding effects." They also suggested that (2) the PLoS study's finding that gold OA generated more citations than green OA in PNAS pertained to OA in general rather than just to high-profile journals like PNAS (and that perhaps green OA is not even OA!):

> **Eysenbach (2006):** "*[T[he [prior] evidence on the "OA advantage" is controversial. Previous research has based claims of an OA citation advantage mainly on studies looking at the impact of self-archived articles... (which some have argued to be different from open access in the narrower sense)... All these previous studies are cross-sectional and are subject to numerous limitations... Limited or no evidence is available on the citation impact of articles originally published as OA that are not confounded by the various biases and additional advantages [?] of self-archiving or "being online" that contribute to the previously observed OA effects*."

> **PLoS Editorial (MacCallum & Parthasarathy 2006):** "*We have long argued that papers freely available in a journal will be more often read and cited than those behind a subscription barrier. However, solid evidence to support or refute such a claim has been surprisingly hard to find. Since most open-access journals are new, comparisons of the effects of open access with established subscription-based journals are easily confounded by age and reputation... As far as we are aware, no other study has compared OA and non-OA articles from the same journal and controlled for so many potentially confounding factors... The results... are clear: in the 4 to 16 months following publication, OA articles gained a significant citation advantage over non-OA articles during the same period... [Eysenbach's] analysis [also] revealed that self-archived articles are... cited less often than OA [sic] articles from the same journal*."

When I pointed out in a reply that subject areas, countries and years *had* all been analyzed separately in prior within-journal comparisons based on far larger samples, always with the same outcome -- the OA citation advantage -- making it highly unlikely that any of the other potentially confounding factors singled out in the PLoS/PNAS study would change that consistent pattern, Eysenbach responded:

> **Eysenbach:** "*[T]o answer Harnad's question* 'What confounding effects does Eysenbach expect from controlling for number of authors in a sample of over a million articles across a dozen disciplines and a dozen years all showing the very same, sizeable OA advantage? Does he seriously think that partialling out the variance in the number of authors would make a dent in that huge, consistent effect?' *– the answer is "absolutely*".

My doctoral student, Chawki Hajjem, has accordingly accepted Eysenbach's challenge, and done the requisite multiple regression analyses, testing not only (3) number of authors, but (1) number of years since publication, and (2) journal impact factor. The outcome is that (4) the OA self-archiving advantage (green OA) continues to be present as a robust, independent, statistically significant factor, alongside factors (1)-(3):

**Tested:**
**(1)** number of years since publication (BLUE)
**(2)** journal impact factor (additional variable not tested by Eysenbach) (PURPLE)
**(3)** number of authors (RED)
**(4)** OA self-archiving (GREEN)

**Already tested separately and confirmed:**
(5) country (previously tested: OAA separately confirmed for all countries tested -- 1st author affiliation)
(6) subject area (previously tested: OAA separately confirmed in all subject areas tested)

**Not tested:**
(7) publication and citation counts for first and last authors (not tested, but see Moed 2006)

**Irrelevant:**
(8) article type (only relevant to PNAS sample)
(9) submission track (only relevant to PNAS sample)
(10) funding type (irrelevant)

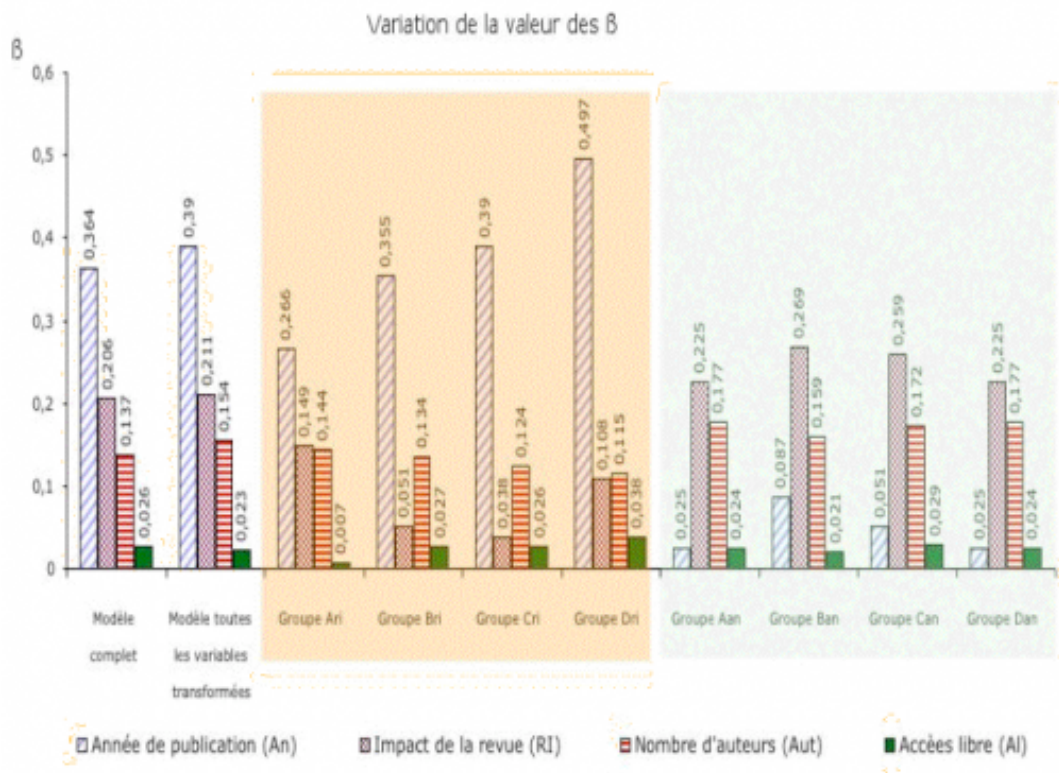

**Independent effects of (1) Year of Publication (purple), (2) Journal Impact Factor (blue), (3) Number of Authors (red) and (4) OA Self-Archiving (green)**

**on citation counts:** Beta weights derived from multiple regression analyses of (column 1) raw distribution, (column 2) log normalized distribution, (columns 3-6) separate Journal Impact Factor Quartiles, and (columns 7-10) separate Year of Publication Quartiles. In every case, OA Self-Archiving makes an independent, statistically significant contribution (highest for the most highly cited articles, column 6 "Groupe Dri": i.e., the QA/QB effect). (Biology, 1992-2003; 576 journals; 442,750 articles). For more details see Chawki Hajjem's website.

In order of size of contribution:

*Article age* (1) is of course the biggest factor: Articles' total citation counts grow as time goes by.

*Journal impact factor* (2) is next: Articles in high-citation journals have higher citation counts: This is not just a circular effect of the fact that journal citation counts are just average journal-article citation counts: It is a *true* QB selection effect (nothing to do with OA!), namely, the higher quality articles tend to be submitted to and selected by the higher quality journals!.

The next contributor to citation counts is the *number of authors* (3): This could be because there are more self-citations when there are more authors; or it could indicate that multi-authored articles tend to be of higher quality.

But last, we have the contribution of *OA self-archiving* (4). It is the smallest of the four factors, but that is unsurprising, as surely article age and quality are the two biggest determinants of citations, whether the articles are OA or non-OA. (Perhaps self-citations are the third biggest contributor). But the OA citation advantage is present for those self-archived articles (and stronger for the higher quality ones, QA), refuting Eysenbach's claim that the green OA advantage is merely the result of "potential confounds" and that only the gold OA advantage is real.

I might add that the PLoS Editorial is quite right to say: "*Since most open-access journals are new, comparisons of the effects of open access with established subscription-based journals are easily confounded by age and reputation*": Comparability and confounding are indeed major problems for *between-journal* comparisons, comparing OA and non-OA journals (gold OA). Until Eysenbach's within-journal PNAS study, "solid evidence" (*for gold OA*) was indeed hard to find. But comparability and confounding are far less of a problem for the *within-journal* analyses of self-archiving (green OA), and with them, solid evidence abounds.

I might further add that the solid pre-existing evidence for the green OA advantage -- free of the limitations of between-journal comparisons -- is and always has been, by the same token, evidence for the gold OA advantage too, for it would be rather foolish and arbitrary to argue that free accessibility is only advantageous to self-archived articles, and not to articles published in OA journals!

Yet that is precisely the kind of generalization Eysenbach seems to want to make (in the opposite direction) in the special case of PNAS -- a very selective, high-profile, high-impact journal. PNAS articles that are freely accessible on the PNAS website were found to have a greater OA advantage than PNAS articles freely accessible only on the author's website. With just a little reflection, however, it is obvious that the most likely reason for this effect is the high profile of PNAS and its website: That effect is hence highly unlikely to scale to all, most, or even many journals; nor is it likely to scale in time, for as green OA grows, the green OA harvesters like OAIster (or even just Google Scholar) will become the natural way and place to search, not the journal's website.

Having taken up Eysenbach's challenge to test the independence of the OA self-archiving advantage from "potential confounds," we now challenge Eysenbach to *test the generality of the PNAS gold/green advantage across the full quality hierarchy of journals, to show it is not merely a high-end effect.*

Let me close by mentioning one variable that Eysenbach did not (and could not) control for, namely, author self-selection bias (Quality Bias, QB): His 212 OA authors were asked to rate the relative urgency, importance, and quality of their articles and there was no difference between their OA and non-OA articles in these self-ratings. But (although I myself am quite ready to agree that there was little or no Quality Bias involved in determining which PNAS authors chose which PNAS articles to make OA gold), unfortunately these self-ratings are not likely to be enough to convince the sceptics who interpret the OA advantage as a Quality Bias (a self-selective tendency to provide OA to higher quality articles) rather than a Quality Advantage (QA) that increases the citations of higher quality articles. Not even the prior evidence of a correlation between earlier downloads and later citations is enough. The positive result of a more objective test of Quality Bias (QB) vs. Quality Advantage (QA) (comparing self-selected vs. mandated self-archiving, and likewise conducted by Chawki Hajjem) will be reported shortly.

**Stevan Harnad & Chawki Hajjem**
American Scientist Open Access Forum